\documentstyle[eqsecnum,aps]{revtex}

\begin{document}
\draft
\preprint{PUC-Rio, UFRJ}
\twocolumn[\hsize\textwidth\columnwidth\hsize\csname
@twocolumnfalse\endcsname
\title{   Magneto-Roton Modes in the Ultra Quantum Crystal: Life and Death at Large Magnetic Fields}
 \author{Pascal Lederer$^+$\footnote{On leave from
Physique des Solides,U. P. S., F91405 Orsay, France( Laboratoire associ\'e au CNRS)} 
 and C. M. Chaves$^{\diamond}$ } 
\address{ $^+$ $^{\diamond}$Depto de Fisica, PUC-Rio, C. P. 38071, Rio de Janeiro  and $^+$Instituto de Fisica, Universidade Federal do Rio de Janeiro, Ilha do Fundao, Rio de Janeiro}
\date{Sep. 18, 1997}
\maketitle
\begin{abstract}
 The Ultra Quantum Crystal phases observed in quasi-one-dimensional conductors of the Bechgaard salts  family under magnetic field exhibit both Spin Density Wave order and a Quantized Hall Effect, which  may exhibit  sign reversals.  We examine the case with no sign reversals.   As the field increases, Magneto-Roton modes , which have minima at quantized wave vectors, merge, or not, in the single particle excitation continuum. In the high field phase,  the Hall conductivity at zero temperature is zero; at the transition from the $N=1$ phase, it has  two Magneto-Roton modes, the energies of which increase rapidly with  magnetic field. The modes  survive until a critical field above which the high field phase  looks more like a perfect nesting insulating Spin Density Wave.
\end{abstract}
\vskip2pc]

\section{ INTRODUCTION}.
Organic conductors of the Bechgaard salts family, $(TMTSF)_2 X$ where $TMTSF$ =
tetramethylselenafulvalene are  quasi-one-dimensional (quasi-1D) systems, which have been found over the last few years to exhibit fascinating properties under magnetic field \cite{review,gorkov,gm91,pl96,hlm1}. The typical hierarchy of their transfer integrals is: $t_a=3000K$,  $t_b=300K$, $t_c=10K$.
In three members of this family ($X=ClO_4,PF_6,ReO_4$), the metallic phase is destroyed by a moderate magnetic field $H$ applied along the $c$ direction, perpendicular to the most conducting planes ($a,b$). A cascade of  magnetic phases, separated by first order transitions appears as the field intensity is stepped up: within each sub-phase, which  are called "Ultra Quantum Crystal" (UQC in the following)\cite{pl87}, a Field Induced Spin-Density Wave  Phase  (FISDW) 
is stabilized  with a peculiar electronic structure, characterized by a small number of exactly filled Landau levels (bands in fact) \cite{pl96}. The Landau bands are separated  by a hierarchy  of gaps $\delta_n$ which oscillate with the magnetic field\cite{phml}. The phase labelled by $N$ has a Hall conductivity $\sigma_{xy}=2Ne^2/h$,
and is characterized by the largest gap $\delta_N$ at the the Fermi level. A sum rule $\Sigma_n \delta_n^2= \Delta^2$, connects all the gaps to the order parameter
$\Delta$. Thus
  each UQC sub-phase exhibits a Quantized Hall conductivity, which is the first example of a Quantum Hall Effect in a 3D system.
This cascade of quantized phases result from an interplay between the nesting properties of the Fermi Surface (FS), and the quantization of 
electronic orbits in the field: the wave vector of the FISDW varies with field so that unpaired carriers in a subphase are always organized in
 completed filled Landau bands. As a result, the number of carriers in each subphase is quantized, and so is the Hall conductivity\cite{review,pl96}. The condensation of the UQC phases results from the peculiar electronic structure of open Fermi Surface metal under magnetic field: because of the Lorentz force, the electronic motion becomes periodic  and confined along the high conductivity direction of the chains ($a$ direction)\cite{gorkov}.
 The periodic motion of the electrons in real space
 is characterized by a wave vector $G=eHb/\hbar$, and an orbital energy $\hbar \omega_c=v_FG/2$, $b$ being the interchain distance. (In the rest of this Letter, wave vectors will be expressed in units of $G$).

 As a result, the static 
bare susceptibility of the normal 
phase, $\chi _0({\bf Q})$ can be expressed 
as a sum over weighted
 strictly 1D bare susceptibilities which diverge at quantized
  values of the longitudinal component of the wave vector $Q^n_{\vert  \vert}=2k_F+n$\cite{gm91,pl96,hlm1}.
 The largest divergence signals the appearance of a SDW phase with quantized    
 vector $Q_{\vert  \vert}=2k_F+N$. This  Quantized Nesting Model (QNM)  \cite{hlm1} describes most of the features of the phase diagram in a magnetic field. It
 has been shown recently to explain  the experimental
 observation of  the Hall plateaux sign reversal when the field varies\cite{zm,rib2,bali}.
 Most plateaux exhibit the same sign. (By convention we will refer to these plateaux
 as positive ones).

 It has been pointed out some years ago \cite{pl87,dppl} that FISDW were a novel kind of electron-hole condensate, an Ultra Quantum Crystal: their condensation energy is smaller than the orbital energy $\hbar \omega_c=\hbar v_FG/2$. As such, it exhibits specific collective modes with a Magneto-Roton (hereafter MR)
structure which has its minimum at a value of the parallel component ($q_{\vert \vert}$) of its momentum equal to  $nG$ ($n$ integer). The original paper on the UQC collective modes suggested that  MR modes might exist at
$q_{\vert \vert}=2G,3G$, etc., but only the mode at $G$ was proved to exist\cite{pl87,dppl}. 

Recently one of us (PL) remarked that the UQC with  "negative " Quantum Hall Number, the "Ribault Phase", exhibits a MR spectrum with no analog in Condensed Matter Physics, {\bf different} from that of the majority sign UQC phases\cite{pl97}. Indeed, when the sequence of observed Quantum Hall Numbers\cite{bali} is : $N=1, 2, -2, 3, 4, 5, 6,7$,  the Ribault Phase (i. e. $N=-2$) has MR modes with $q_{\vert \vert}=4$ or $5$ well within the single particle gap, and no low energy mode at $q_{\vert \vert}=1$ or $2$. It was shown in \cite{pl97} that the Ribault phase "contaminates" the neighbouring phases in parameter space, as well as those thermodynamically close by. In the latter example, the phase  $N=2$ (resp $N=3$) has at least one additional mode at $q_{\vert \vert}=4$ (resp. $q_{\vert \vert}=5$). Examples of Hall sequences with more sign changes
have been shown to have more structure yet in the collective modes, with possibly $4$, etc., MR  modes within the single particle gap.
 Furthermore, in contrast to the "usual" MR, the two MR in the Ribault phase have large monotonic variations, with opposite signs, as the field varies\cite{pl97}.

The latter findings prompted us to reexamine in more details, and in a quantitative way, the MR of the usual UQC, when no sign change of the Hall Effect occurs as the field varies\cite{bali}.

This Letter gives for the first time  numerical estimates, based on the RPA, of the MR modes
energies in $(TMTSF)_2X$,  of their magnetic field dependence, and of the MR transverse effective mass. We find a distinctly different behaviour of the UQC at high fields ($N=1$ or $0$)  from that at low field $N\geq  2$.

\section{  MR IN THE GENERIC CASE}
We are interested in the usual UQC with no sign inversion of the Quantum Hall Effect. Therefore it is sufficient to consider the following electronic dispersion relation:

\begin{eqnarray} \label{model}
\epsilon({\bf k})& =& v_F(\vert k_x\vert - k_F)+ \epsilon_{\perp }({\bf k_{\perp}}),\\
\epsilon_{\perp }({\bf k_{\perp}})& =& -2t_b\cos k_yb  -2t_c\cos k_zc -
2 t'_b\cos 2k_yb 
 \nonumber
\end{eqnarray}
$\epsilon(k_{\perp})$ is a  periodic function which describes a warped FS.

The  normal metal-FISDW instability line  $T_{cN}(H)$ is given by: 

\begin{equation} \label{ki}
\chi _0({\bf Q},T_{cN}, H)=
\Sigma _nI_n^2 (Q_{\perp })\chi _0^{1D}(Q_{\vert  \vert}-n, T_{cN})=1/\lambda
\end{equation}

$\lambda$ is the electronic interaction constant. Eq.(\ref{ki})
 exhibits the structure of $\chi _0$ as the sum of one
 dimensional terms $\chi _0^{1D}$ shifted by the magnetic field wave
 vector $G=eHb/\hbar$. 
$\chi_0^{1D}\propto -\ln(\max \{ v_F(2k_F-q),T   \}/\epsilon_F)$. 
In eq.(\ref{ki}),  the coefficient $I_n$ depends on the dispersion  relation and H:
\begin{eqnarray}
I_n(Q_{\perp}) &   =& \overline{
 \exp i\left[(T_{\perp}(p+Q_{\perp}/2)+ 
  T_{\perp}(p-Q_{\perp}/2)+np \right] }
\end{eqnarray}
where $T_{\perp}(p) = (1/\hbar \omega _c)\int_0^p\epsilon_{\perp}( p')dp'$
and $\overline{A(p)}$ denotes the average of $A$ over p.

Define $T^*$ as the FISDW ordering temperature in infinite field. From eq.(\ref{ki}) ,
we have $T^*=(2\gamma/\pi)E_0 \exp(-1/\lambda)$. ($\gamma$ is Euler's constant and $E_0$ a high energy cut-off). $T^*$ is equal to the ordering temperature for the perfectly nested Fermi Surface at any field (i. e. when $t'_b=0$). Now define a generalized instability temperature $T_{N \pm m}(\pm q_{\perp})$:

\begin{equation} \label{Tc}
   \frac{ T_{N \pm m}}{T^*}     = \exp \left[                  
\sum_{n \neq  0} \frac{ I^2_{N\pm m +n} (Q_{\perp }^N \pm q_{\perp})}{  I^2_{N\pm m}(Q^N_{\perp } \pm q_{\perp }) }\ln \left(    \frac{\pi T^*}{2\gamma \vert n\vert \omega _c}\right) \right]
\end{equation}
In (\ref{Tc}), we have used the sum rule: $\Sigma_n I_n^2=1$.
For $m=0$ and $q_{\perp}=0$, $T_{N \pm m}=T_{cN}$, the ordering
 temperature for the $N$th subphase. For $m\neq 0$, $T_{N \pm m}(q_{\perp})$ generalizes the definition of the critical temperatures on either side
 of phase $N$ in the $(T,H)$ plane. $T_{N \pm m}(q_{\perp})$ are at
 most equal to the virtual transition lines $T_{ N\pm m}$ which can 
be drawn in the $N$th subphase part of the phase diagram and which 
represent virtual transition lines to phases with slightly larger
 free energy than the $N$th subphase\cite{pl87}. In the ($T,H$)
 plane, there is an infinite number of continuous lines  crossing 
the phase diagram. The upper limit of this family is the actual
 (continuous non analytic) transition line from the normal metal
 to the UQC; this  line coincides piecewise with the transition lines
 labelled by  the successive integers describing the Quantum Hall
conductivity. See Fig.(1) .
 An example of computed network of 
transition lines was given in \cite{lm}

MR energies have  been derived within the RPA\cite{pl87,dppl},
 by looking at the poles of the spin-spin correlation function
 of the ordered phase. The electronic orbital motion, together with the RPA treatments of interactions,  results in 
effective scattering potential energy terms which
couple electron states with wave vector $k_{\vert \vert}$
 and $k_{\vert \vert}+2k_F +n$, (n integer) on either side 
of the Fermi Surface. This results in a series of gaps in 
the condensed phase dispersion relation\cite{phml}, corresponding
 to the various potential scattering terms. The simplest 
approximation which captures the esssential physics resums to 
all orders the gap $\delta_N= I_N \times  \Delta$ at the Fermi level
 and takes all other gaps into account to second order in perturbation\cite{pl87,dppl}. The limit $2\delta_n/\omega_c\ll 1$
 is assumed to hold; in fact this ratio is small at $H\lesssim 4$T, and is small again at large  $H$, since $2\delta^*=3.53\times T^*$ 
is a fixed bound.   Then the  equation for collective modes 
in the UQC phase $N$ reduces to:

\begin{eqnarray} \label{rot2}
\left( \ln \left(  \frac{2\gamma E_0}{\pi T_{N+m }} \right) -   \tilde{\bar{\chi_0}}(\delta,\omega)                                             \right) && \nonumber \\
 \left(  \ln \left(  \frac{2\gamma E_0}{\pi T_{N-m}} \right) -\tilde{\bar{\chi_0}}(\delta, \omega)   \right) &  
=&\left( \tilde{\bar {\Gamma_0}} (\delta ,\omega )    \right)^2 
\end{eqnarray}

where $T_{N \pm m}$ is defined in eq.( \ref{Tc}), and 
$q_{\vert  \vert}=2k_F +m +\delta$, with $ \delta<<1$
.
$\tilde{\bar{\chi_0}}$ and $\tilde{\bar{\Gamma_0}}$ are for $n=0$ the
 objects discussed in \cite{lra} in connections with collective modes of SDW. 
Eq.(\ref{rot2}) finally yields, setting $x_{N,m}=\omega(q_{\vert \vert}=m+\delta, q_\perp)/(2\delta_N)$, with ($q_z=\pi/c$):
\begin{eqnarray}  \label{Tm}
(2x_{N,m}^2-1)h(x_{N,m})&=&\ln \left( \frac{T_{cN}^2}{T_{N+m}T_{N-m}}\right)
\nonumber \\
+ \left( \ln^2 \left( \frac{T_{N-m}}{T_{N+m}}\right) +h(x_{N,m})^2                                           \right)^{1/2}  &&
\end{eqnarray}
with
$h(x ) =
\int_0^\infty du\frac{\tanh (\frac{\delta_N}{2T}\cosh u)}{\cosh^2u-x ^2}$. 
We have computed MR energies at $0$K  in the generic UQC phase labelled by $N=4$
 from eqs.(\ref{Tc}), varying $q_{\perp}$ in order to get the 
locus of the various MR minima in the $q_\perp $ direction. As a by-product, 
we obtain for the first time a numerical estimate of the transverse
 effective mass $m_\perp$ defined in phase $N$ by $\omega _N(q_\perp, q_{\vert \vert}=m)= \omega^{min}_{N,m}+q_\perp^2/2m_\perp$ ($\hbar=1$).
 The parameters we took for the numerical estimates are relevant to the Bechgaard salts, and have been used previously in the literature\cite{lm}: $t_a=3000K$, $t_b=265K$, $t'_b=10K$, $T^*=6.44$. 
(Those parameters may change with pressure, notably $t'_b$).
 The results are  shown on fig.(2). We find two modes, at $q_{\vert \vert}=1$ or $2$ which vary little with $H$,
and a {\bf third mode}, rather close to the upper edge of the gap, at {\bf $q_{\vert \vert }=6, q_\perp=0$}. This mode is yet another example of "contamination" by the Ribault phase $N=-2$ discussed in ref. \cite{pl97}. It has a monotonic $H$ negative variation, and drowns into the single particle continuum for $H\lesssim 3.75$T.
 We have looked for MR mode with $q_{\vert \vert}=8$, in phase $4$, which might arise from contamination of the near-by Ribault phase $N=-4$, and found it did not exist in the absence of additional terms in eq.(\ref{model}), but this mode will certainly become alive under pressure\cite{zm,pl97}.

 The RPA evaluation yields
 a MR minimum $x_{5,1}^{min}\simeq .4$, located at $q_\perp=.0178/b$
 for $H=3.6T$. We find  a {\bf very large} mass anisotropy 
 $m_{\perp}/m_{\vert \vert}|_{N=5,m=1}\simeq .9 \times 10^3$, with
 $m_{\vert \vert}= \hbar \omega_{min}/v_F^2$. The fact that
 $x_{5,1}<1/2$ makes this MR mode a good candidate for the
 kind of specific heat anomalies discussed in \cite{pl96,pl97b}.
 The mass anisotropy is larger for the mode $q_{\vert \vert}=2$,
 we find $\simeq$ $2.25\times 10^3$.See fig.(3). Note that the MR energy $\omega_{Nm}^{min}$ is discontinuous at the transition from $N$ to
 $N\pm 1$. This contributes (a small term, which corrects the mean field estimate) to the transition latent heat\cite{pl97b}.  
 On the other hand, $\omega^{min}_{4,6}$ goes continuously at the transition to $N=3$ into $\omega^{min}_{3,5}$ (i. e. a {\bf distinct} MR with the {\bf same} energy!) 

\section{ LIFE AND DEATH AT LARGE MAGNETIC FIELD}
The situation changes qualitatively as soon as the magnetic 
field exceeds $6.1$ T, the critical value for the transition 
from UQC phases $2$ to $1$. The results are shown on fig.(4). 
The two MR which exist in the low field part of phase $N=1$ have 
different field dependences. The high energy mode, with $q_{\vert \vert}=2$ increases its energy monotonically with field, at a rate $\simeq .6K/T $, 
then merges into the single particle continuum for $H \gtrsim7$ T.
 Then only one MR at $q_{\vert \vert}=1$ survives, at lower energy, 
and with a non monotonic field dependence. In fact the rate of variation of $\omega^{min}_{1,1}$ is also large,
$\simeq -1.8K/T$ for $H\gtrsim 7.5$ T. For $H\lesssim 6.5$ T, $d\omega^{min}_{1,1}/dH\simeq .4K/T$.

At the transition from $N=1$ to $0$, at $H\simeq 8.25$ T,
 two MR 
modes appear discontinuously. The MR at $q_{\vert \vert}=1$ is
 $21 \%$ 
higher in energy in phase $0$, and that at $q_{\vert \vert}=2$ is
 close
 to the single particle continuum, at $.95 \times 2\delta_0$. In fact 
this latter mode is also connected to the "contamination" of the Ribault
phase at $N=-2$, since it connects $T_2$ and $T_{-2}$ through eq. (\ref{Tc})\cite{pl97}.
 Both MR energies increase rapidly with $H$, at a rate $\simeq .5 K/T$ 
for the low energy mode, $.8 K/T$ for the high energy one.       
  The last surviving MR dissolves into the single particle continuum for $H\gtrsim 12.5$ T. For larger $H$ the only collective modes left
 are the usual Goldstone modes at $q_{\vert \vert }\ll 1$. Then  the only specific features of the UQC left are small single particle energy gaps, at most an order of magnitude smaller than the FS gap, at large energies
 $\hbar \omega_c,\;
 2\hbar \omega_c$, etc.. The low energy physics of the $N=0$ phase is identically that of an insulating SDW phase with perfect nesting FS.

The actual field values for which the MR mode drowns in the
 continuum in phase $N=0$ cannot be taken too seriously, since for that 
range of field values, the approximation $2\delta_0/\hbar \omega_c \ll 1$ breaks down completely. However, there is little doubt that the behaviour found  above
holds qualitatively since the approximation becomes good again at higher fields.  

\section{ CONCLUSION}

In addition to proving that within the RPA, more than one MR mode is usually alive in the single particle gap, at $q_{\vert \vert}=1$ or $2$ (possibly $3$, or $5,6, 8,$ etc. due to the vicinity of Ribault phases) and computing their energies as a function of field, we have derived for the first time numerical estimates of the very large effective mass anisotropy of the MR. We have described how, as the magnetic field increases, UQC phases at small quantum numbers progressively lose their MR and, above a threshhold field, retrieve a low energy excitation spectrum of a conventional SDW with perfect nesting.
Our results are obtained within the RPA, in the "very weak" coupling limit ($2\delta_N / \hbar \omega_c \ll 1$). In Bechgaard salts, this ratio is of order .4 below 6 T, and increases with field. Between 8 T and 20 T, it is actually larger than 1, then it decreases and becomes small at very large fields.  Thus, significant corrections may be expected, in a weak coupling RPA, (i. e. $\lambda \ll 1$) to all estimates derived here, in particular between 8 T and 20 T. We have found that the lowest MR energy minima of the positive UQC phases could be sometimes below $\delta_N$. Those values might be significantly lowered by self-trapping effects\cite{hlm}, which are outside the scope of linear response theory\cite{pl97b}. However the presence of MR with $x_{N,m}< .5$ does not seem to be garanteed for all $N$. The effective masses should increase due to MR-MR scattering processes,etc.. The importance of those effects will be gauged by comparing to the estimates given here. The latter should be a useful guideline for experimental efforts to
determine the MR parameters.

\acknowledgements
We would like to thank Daniel Frenkel for his help with the Maple software we have been using. One of us (PL) thanks the CNPq for financial support during completion of this work, and Gilson Carneiro, for his hospitality in UFRJ.

\begin{figure}
\caption{Network of $T_{N\pm m}(q_\perp)$ lines in the $(T,H)$ plane for $6$T$<H<12.5$ T. Dotted lines are the upper limit of the family obtained by varying $q_\perp $. At $T=0 K$,  $2\delta_N\simeq 3.53 \times T_{cN}$}
\end{figure}
\begin{figure}
\caption{MR  energy minima as function of H for the generic UQC phase $N=4$ and neighbours. The lowest energy MR is for $q_{\vert \vert}=1$, the second lowest one for $2$. Both are almost H independent. Both modes are discontinuous at the transition from sub-phase to sub-phase. The MR with $q_{\vert \vert}=6$, due to contamination from the Ribault phase $-2$ lies highest in energy, appears only for $H> 3.75$T and has a monotonic negative variation with $H$. A mode at $q_{\vert \vert}=3$ appears very close to the single particle continuum in phase $5$ at $H\simeq 3.6$}
\end{figure}
\begin{figure}
\caption{Dispersion of the MR modes in the transverse direction for the two low energy modes in UQC phase $5$; results are similar for UQC $4$. The effective mass $m_\perp$ is found to be about $10^3$ larger than $m_{\vert \vert}$. }
\end{figure}
\begin{figure}
\caption{MR energy minima as function of H for the high field UQC phases $N=1$ and $N=0$. The phase $1$ loses its high energy MR above $H=7.5$T. The phase $0$ loses its MR one by one as the field increases. Above $H\simeq 12.5$T, the only collective modes left are the Goldstone bososns connected with the SDW  broken symmetries}
\end{figure}
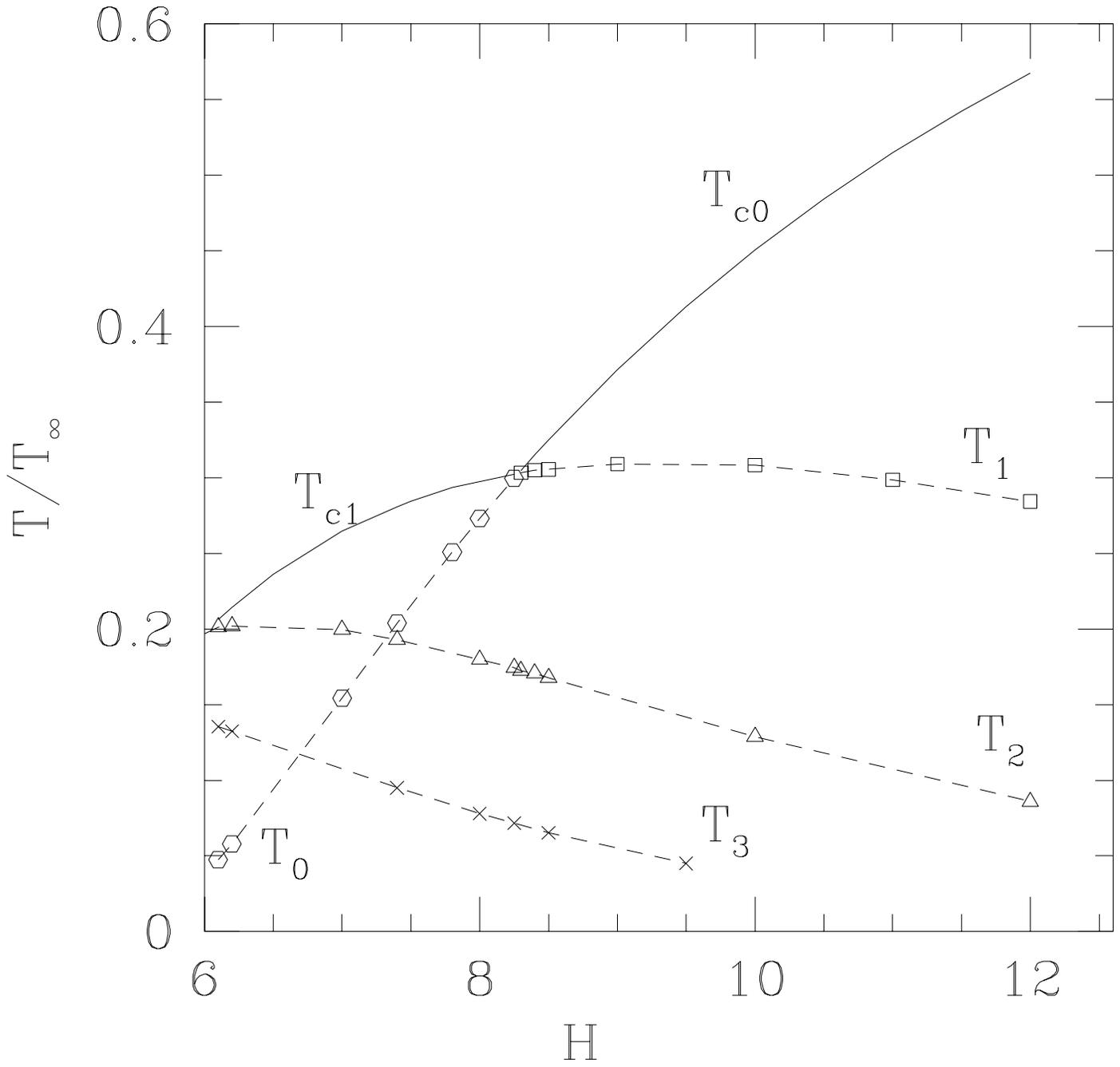
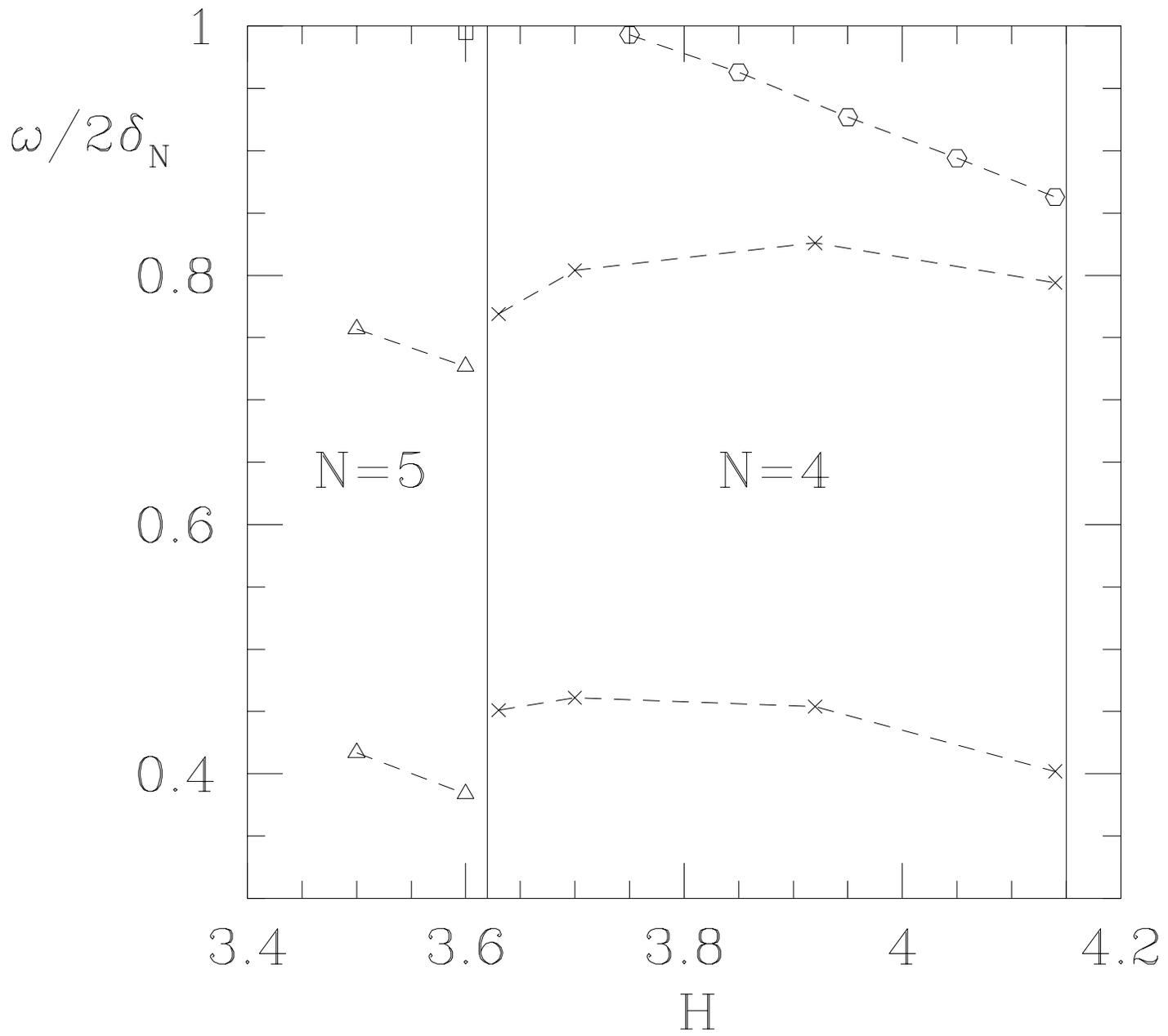
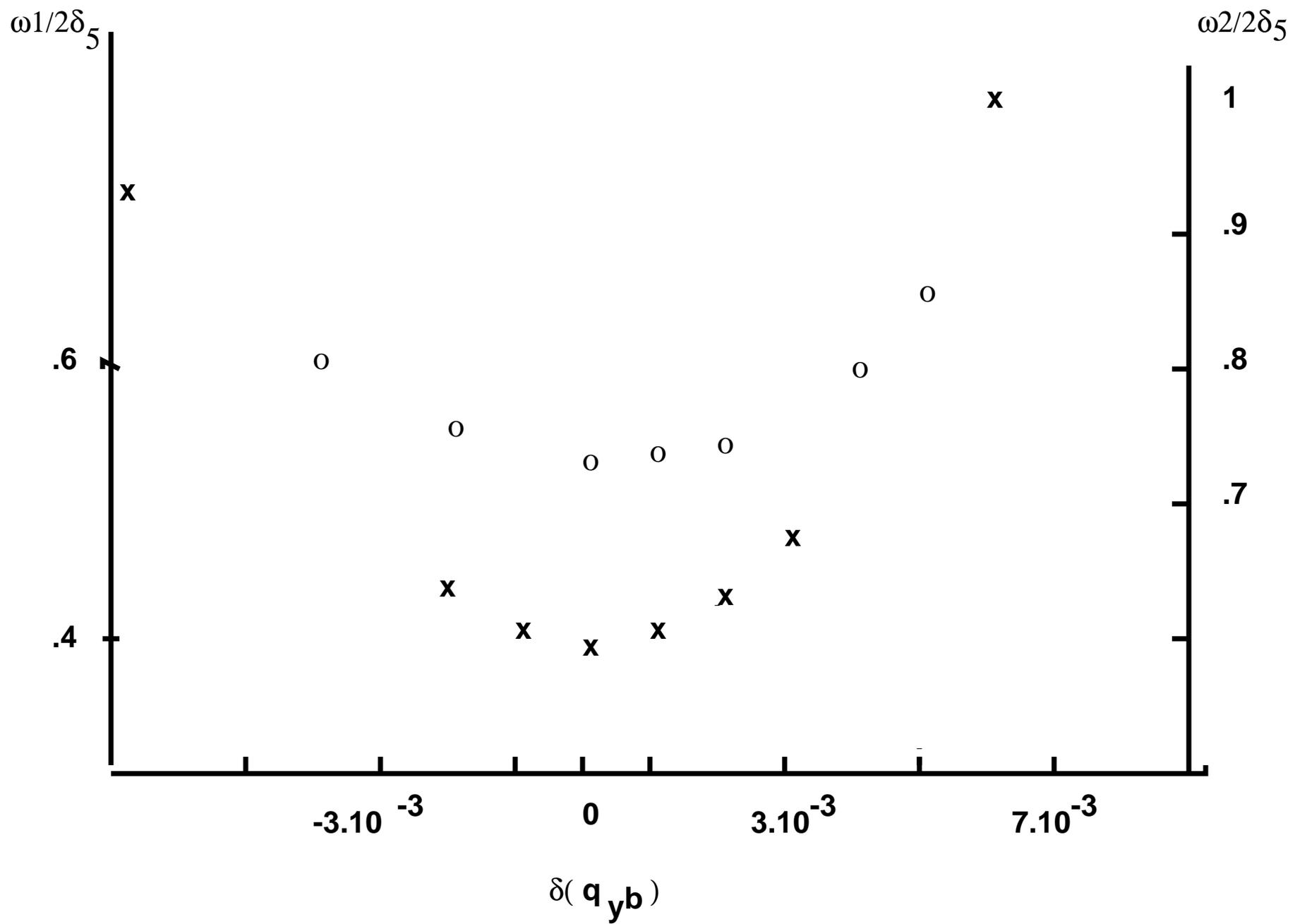
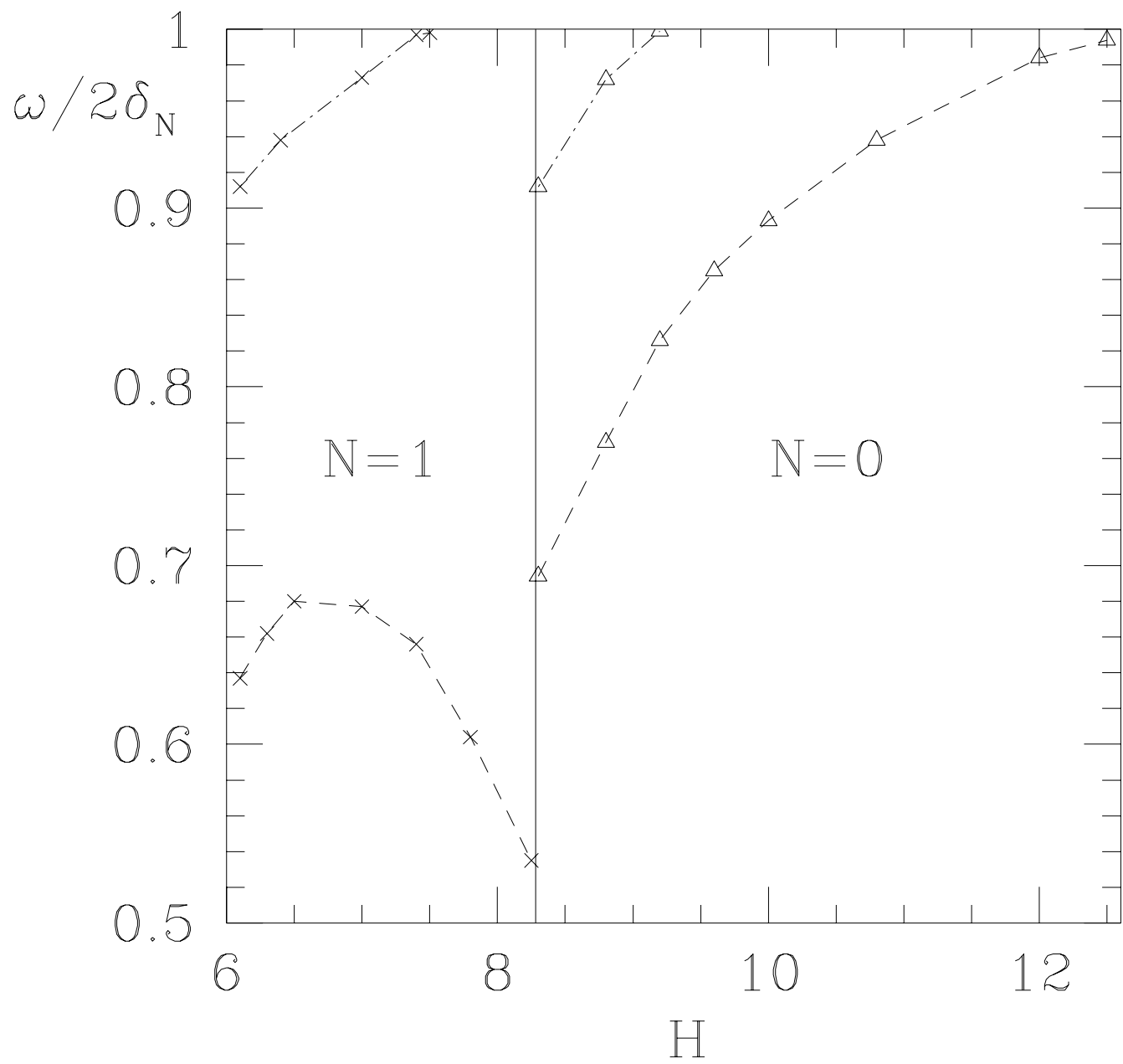
\end{document}